\documentclass[%
reprint,
superscriptaddress,
amssymb,
aps,
prl,
]
{revtex4-1}

\usepackage[english]{babel} 
\usepackage [table]{xcolor}

\usepackage{graphicx} 

\usepackage{amsmath}
\usepackage{amsthm}
\usepackage{amsfonts}
\usepackage{color}

\usepackage{float}
\usepackage{textcomp}
\usepackage{graphicx}
\usepackage{dcolumn}
\usepackage{bm}
\usepackage{hyperref}
\usepackage[mathlines]{lineno}

\begin{document}



\title{Evidence for interacting two-level systems from the \\1/f noise of a superconducting resonator}

\author{J.~Burnett}

\affiliation{ National Physical Laboratory, Hampton Road, Teddington, TW11 0LW,
UK}

\affiliation{ Royal Holloway, University of London, Egham, TW20 0EX, UK}

\author{L. Faoro}

\affiliation{Laboratoire de Physique Theorique et Hautes Energies, CNRS UMR 7589,
Universites Paris 6 et 7, 4 place Jussieu, 75252 Paris, Cedex 05,
France}

\affiliation{Department of Physics and Astronomy, Rutgers The State University
of New Jersey, 136 Frelinghuysen Road, Piscataway, 08854 New Jersey,
USA}

\author{I.~Wisby}

\affiliation{ National Physical Laboratory, Hampton Road, Teddington, TW11 0LW,
UK}

\affiliation{ Royal Holloway, University of London, Egham, TW20 0EX, UK}

\author{V. L.~Gurtovoi}

\author{A.~V.~Chernykh}

\author{G.~M.~Mikhailov}

\author{V.~A.~Tulin}

\affiliation{Institute of Microelectronics Technology and High Purity Materials, Academician Ossipyan str., 6, Chernogolovka, Moscow Region, 142432, Russia}

\author{R.~Shaikhaidarov}

\author{V.~Antonov}

\affiliation{ Royal Holloway, University of London, Egham, TW20 0EX, UK}

\author{P. J.~Meeson}

\affiliation{ Royal Holloway, University of London, Egham, TW20 0EX, UK}

\author{A.~Ya.~Tzalenchuk}
\affiliation{ National Physical Laboratory, Hampton Road, Teddington, TW11 0LW,
UK}
\affiliation{ Royal Holloway, University of London, Egham, TW20 0EX, UK}

\author{T.~Lindstr\"{o}m}

\affiliation{ National Physical Laboratory, Hampton Road, Teddington, TW11 0LW,
UK}

\email{tobias.lindstrom@npl.co.uk}

\selectlanguage{english}%

\date{\today}

\begin{abstract}

Here we find the increase in 1/f noise of superconducting resonators at low temperatures to be completely incompatible with the standard tunnelling model (STM) of Two Level Systems (TLS), which has been used to describe low-frequency noise for over three decades. We revise the STM to include a significant TLS - TLS interaction. The new model is able to explain the temperature and power dependence of noise in our measurements, as well as recent studies of individual TLS lifetimes using superconducting qubits. The measurements were made possible by implementing an ultra-stable frequency-tracking technique used in atomic clocks and  by producing a superconducting resonator insensitive to temperature fluctuations. The latter required an epitaxially grown Nb film with a Pt capping layer to minimize detrimental oxides at all interfaces. 
\end{abstract}

\maketitle

\begin{figure}[!ht]
\centering \includegraphics[ 
width=\columnwidth
]{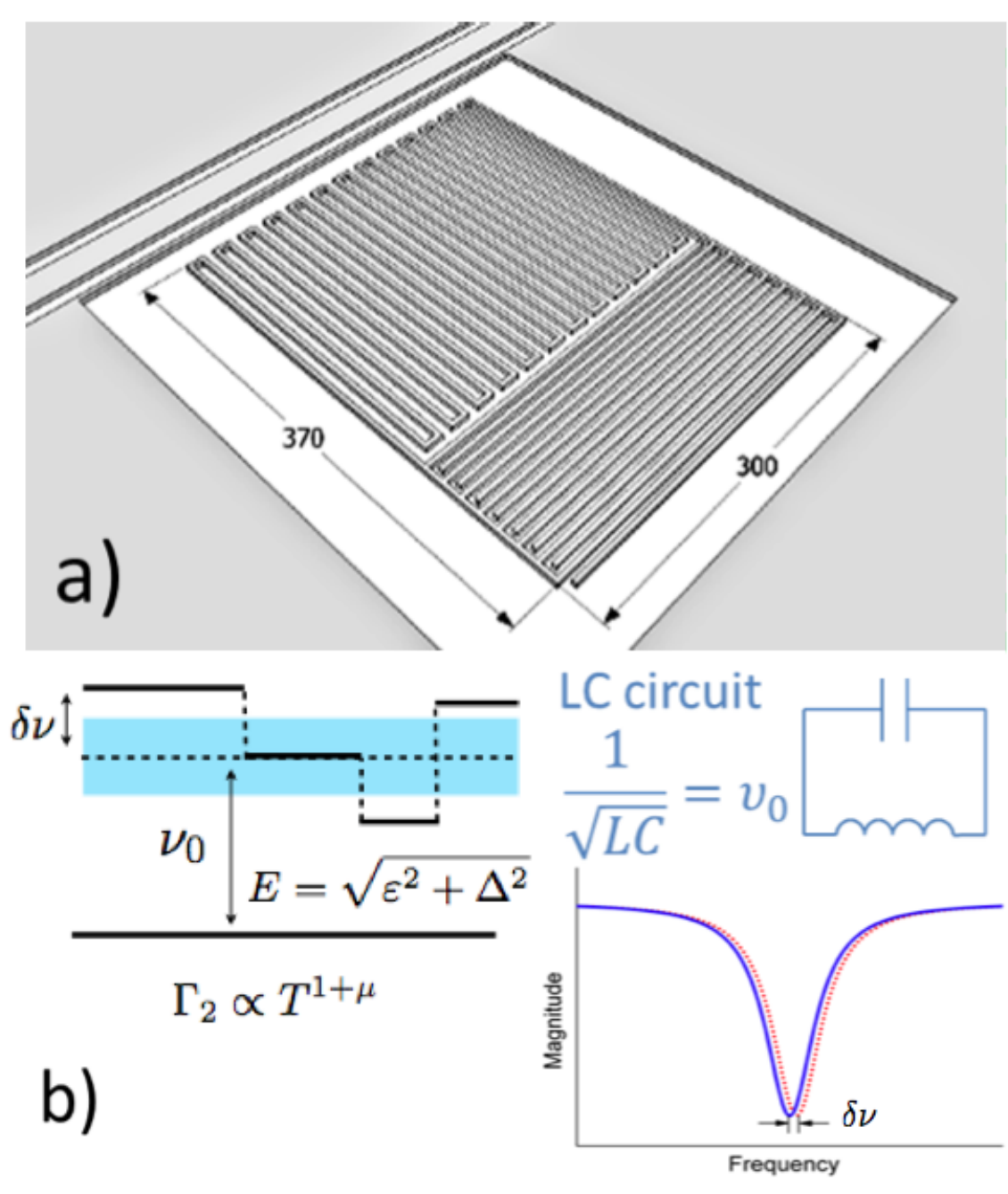}
\caption{\textit{a:} Layout of a superconducting resonator comprised of an
inter-digital capacitor and an inductive meander. The metallization
layer is shown in blue and exposed substrate is shown in white. All dimensions are in $\mu$m and the
lines are 4 $\mu$m wide. The resonator is inductively coupled to a coplanar feedline. \textit{b:} Conceptual model for the noise: quantum coherent TLS with energy splitting ${E \approx \nu_0}$ acquires a width $\Gamma_2 \propto T^{1+\mu}$ due to dipolar dipolar interactions with a bath of TLS having density of states ${\rho(E) \propto E^\mu}$. Here $\mu \approx 0.3$ and $T$ is temperature.  The resonant coherent TLS are in turn affected by surrounding classical fluctuators strongly coupled to them that cause an energy drift, bringing the resonant TLS in and out of resonance. The LC resonator is sensitive to this energy drift and the result is a frequency shift $\delta \nu$, shown in the magnitude response.}
\label{fig:resonator} 
\end{figure}

Initially introduced to explain unusual thermal properties of glasses with respect to crystals, the Standard Tunnelling Model (STM)  \cite{anderson1972,phillips1972}  has enjoyed considerable success in describing many other phenomena in disordered systems aswell. The STM assumes the ability for some atoms at low temperatures to quantum mechanically tunnel between adjacent lattice sites. The model has since been adopted to describe two-level systems (TLS) in many other contexts: for example TLS can be caused by tunnelling electrons in barriers, molecules with two quasi-stable configurations etc.\cite{rogers1986characterization,rella1996vibrational,guimbretiere2012acoustic}. More generally, any influence of dilute impurities on macroscopic properties of materials is often parametrized as due to a bath of non-interacting TLS; so that the STM is applicable. In particular, TLS have been suggested as sources of ubiquitous 1/f noise in electronic materials\cite{black1983nearly} and are also active at low-temperatures, where the kinetics are  dominated by tunnelling \cite{ludviksson1984low,weissman19881}. With the rapid recent progress of new devices for quantum information processing (QIP) and detector applications, there is a renewed awareness \cite{leggett2013tunneling} of the deleterious effects of TLS noise. Here we demonstrate theoretically and experimentally that the STM fails in the extreme low-temperature, low-power limits where these new devices are typically operated and an extension of STM to include specific interactions between TLS is necessary to address the shortcomings of this tried and trusted model.

The properties of TLS can be conveniently  studied using thin-film high quality superconducting microwave resonators, which are important for a number of diverse applications ranging from quantum computation \cite{Wallraff2004} to submillimeter and far-infrared astronomy\cite{day2003}.  Here TLS,  mainly located at interfaces,  cause 1/f noise that manifests as an instability in the centre frequency, and device performance can be severely hampered by the resulting jitter.  The mechanism can be understood by noting that the TLS are associated with dipoles that couple strongly to the electric field and thus modifies the dielectric constant, causing them to be a source  of noise. For this reason there have been a number of studies of TLS noise in superconducting resonators\cite{Gao2007,Gao2008a,Kumar2008,barends2009noise,Barends2010,Burnett2013}, however the results have not been explainable with the STM. This has motivated the development of new models which either assume the presence of additional sources of dissipation e.g. caused by edge defects \cite{neill2013} or alternatively, explain the shortcomings of the STM by re-evaluating the underlying assumptions about the nature of TLS. 

Here we study 1/f noise in specially designed low noise epitaxial superconducting resonators with high temperature stability. We find the results to be inconsistent with the STM. To account for these results we propose a new model, which includes interactions between TLS.  The interactions lead to spectral diffusion which alters the properties of the resonator. 
We show that as well as explaining our results,  the new model  also explains results of other recent studies \cite{lisenfeld2010,sank2012flux} that used phase qubits as a way to probe individual TLS. 

Our model is conceptually shown in fig~\ref{fig:resonator}: a  resonator (fig~\ref{fig:resonator}a) with centre frequency $\nu_0$ couples to an ensemble of coherent TLS with energy splitting $h\nu_0$. Due to non negligible interactions with surrounding TLS, the resonant TLS acquire a temperature dependent width and are affected by strongly coupled slow incoherent (``classical'') fluctuators (fig~\ref{fig:resonator}b). The latter cause an energy drift bringing the resonant TLS in and out of resonance.  One key prediction of this new model is that it results in a low frequency noise for a superconducting resonator that  \textit{increases} as the temperature is decreased. Moreover, this increase can be directly related to a microscopic parameter $\mu$ associated with the slight suppression of the density of TLS states at low energy.

By performing a detailed study and analysis of the very low frequency noise in high quality Nb resonators we are able to validate this model and extract values for $ \mu$.  These measurements are performed in various microwave powers, down to average energies corresponding to ${\left<n\right>=1-10}$ photons in the resonator.  A crucial feature of our measurements is the long measurement time  ($\tau>10^{4}s$) leading to high statistical confidence. 
Two resonators, denoted Res1 and Res2 are studied in detail. The resonators (shown in fig~\ref{fig:resonator}a) are made of 50~nm thick epitaxial Nb film with a 5~nm Pt capping layer grown by laser ablation on r-plane sapphire (see Methods for details). The resonators have centre frequencies ($\nu_{0}$) are between 4-8 GHz, and intrinsic quality factors ($Q$) above $10^{5}$. Further parameters can be found in table \ref{tab:params} of Methods.

The measurements were performed by continuously monitoring the time-varying centre frequency $\nu_{0}(t)$.
For all measurements presented in this work, the quantity of interest is the fractional frequency spectra. ${S_{y}=\left\langle \delta f(t)\delta f(t')\right\rangle /\nu_{0}^{2}}$, where, $\nu_{0}$ is the nominal centre frequency of the resonator, and $f(t)$  the Fourier frequency (see Methods for details).

\begin{figure*}[!ht]
\centering \includegraphics[
]{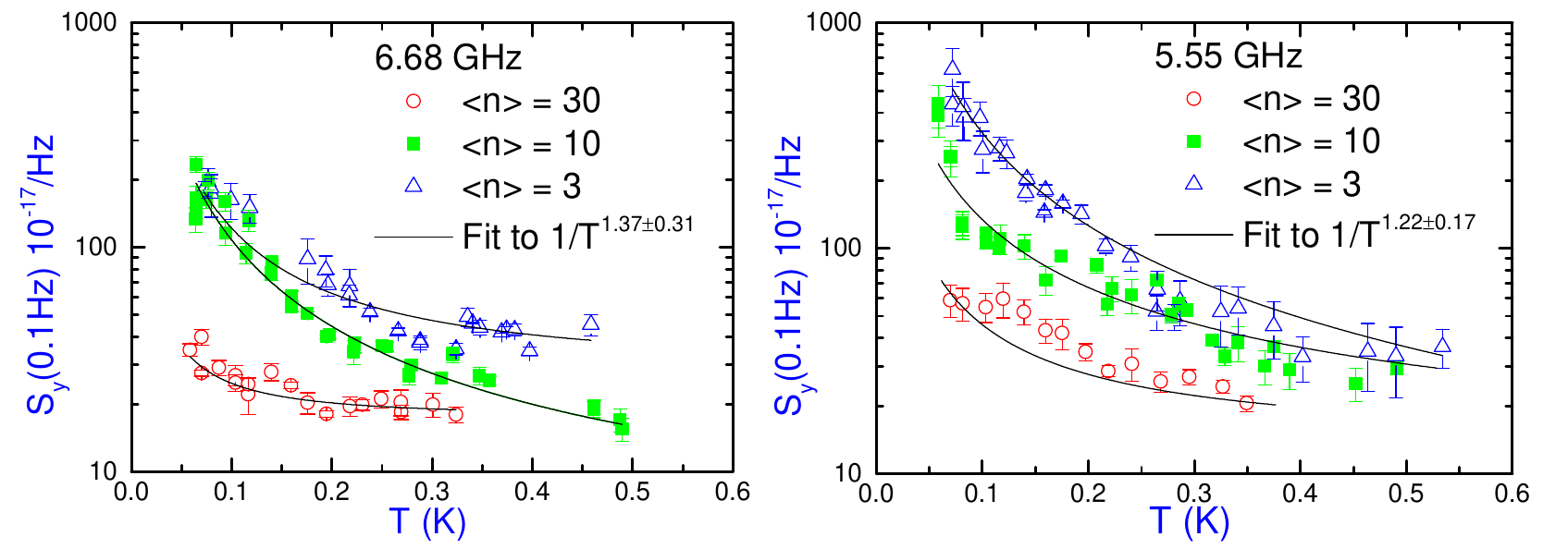}
\caption{ 1/f noise measured at $S_{y}(\text{0}.1~Hz)$ in varying
temperature for different average photon energies in the resonator.
Shown in red is fit to a power law highlighting an inverse temperature
dependence. The noise saturates at a power-dependent level above the
system noise floor of $S_{y}(0.1~Hz)$~=~5x10$^{-17}$. The error
bars indicate type~A uncertainties.}
\label{fig:Tem1} 
\end{figure*}

The central result of this work can be found in fig. \ref{fig:Tem1},
which shows the level of 1/f (fig.~\ref{fig:Tem1}a) noise at 0.1~Hz plotted for the temperature
range 60 mK - 500 mK at three different microwave drive levels (fig.~\ref{fig:Tem1}b). For both resonators we find a stronger than 1/T temperature dependence for all microwave drives. This data can be fit to  $S_{y}(T)\propto a/T^{1+\mu}$, where $a$ is a power dependent constant. Where $\mu=0.22\pm0.16$ and $\mu=0.36\pm0.3$ for Res1 and Res2 respectively.  Hence, both sets of measurement suggest a temperature
dependence stronger than 1/T. 

\begin{figure*}[!ht]
\centering \includegraphics[ 
]{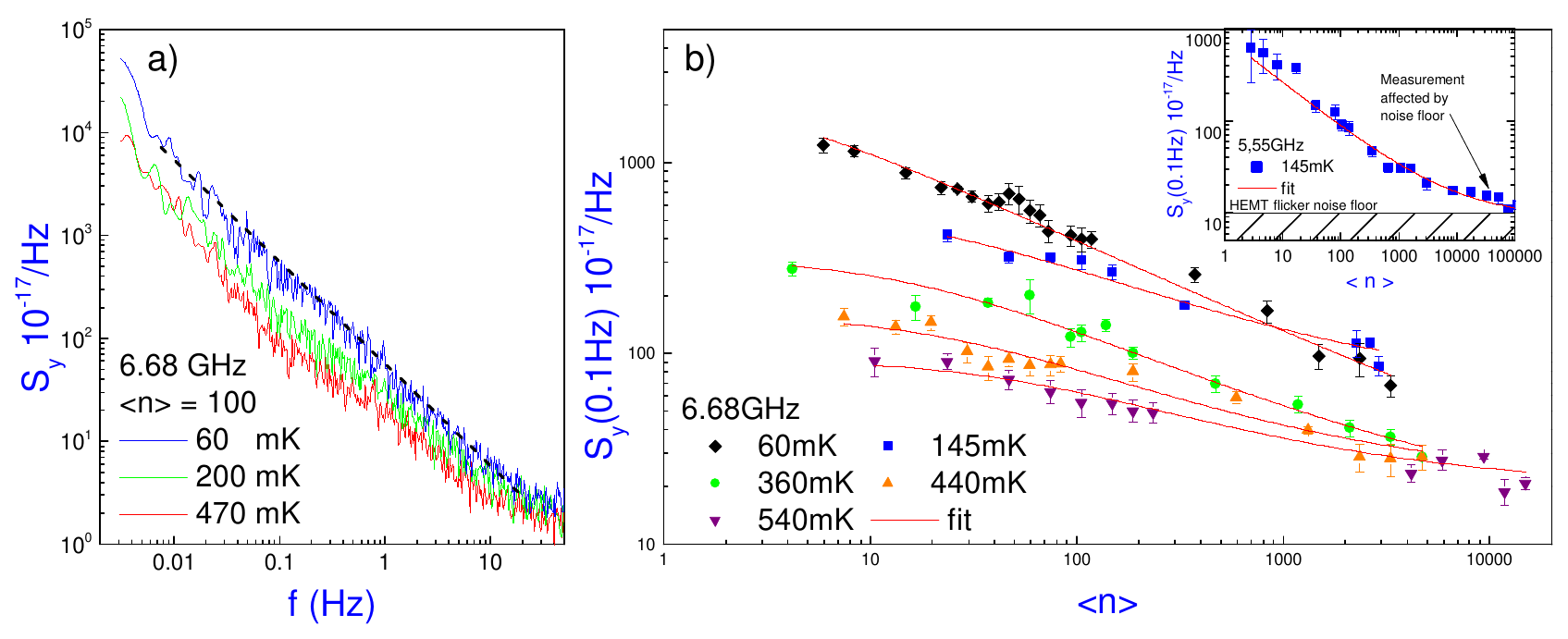}
\caption{\textit{a:} Power spectral density at three different
temperatures for resonator Res2. For frequencies below 10~Hz the slope is -0.97 with a goodness-of-fit parameter $R^{2}=0.9$, white noise dominates from 50~Hz. 
\textit{b:} 1/f noise level vs. average number of photons within the resonator Res2 at several
different temperatures. The average photon number is calculated
by $\left<n\right> = W_{sto}/\hbar\omega_{o}$, where $h$ is Planck's constant, $\omega_{0}$ is the resonators centre frequency and $W_{sto}=2P_{app}S_{min}^{21}Q_{l}/\omega_{0}$. Where $P_{app}$ is the applied microwave drive and $S_{min}$ is the magnitude of the resonance dip. The data is fit to the noise spectra calculated from the model. 
 \textit{inset:} 1/f noise vs. average number of photons for resonator Res1 showing the leveling off at high photon numbers (high microwave drive) that is due to  flicker noise floor of the amplifier chain.
}
\label{fig:PSD_power} 
\end{figure*}

We also find that the 1/f level depends on the microwave
drive, or more precisely, on the energy stored in the resonator; consistent with our previous results\cite{Burnett2013}. At the highest powers (inset of fig~\ref{fig:Tem1}a) we see a saturation due to amplifier noise at a level consistent with other estimates\cite{murch2012}\cite{neill2013}. However, fig~\ref{fig:PSD_power} shows that as power is decreased the resonator noise increases, until it saturates in the low power limit, at a level found to be temperature dependent.

We first discuss whether the jitter in the resonator frequency may be caused by inductance fluctuations from surface spins. The argument in favour
of this is that both frequency jitter and inductance fluctuations
\cite{sendelbach2009} exhibit a $1/f$ frequency
dependence for $T\lesssim200\, mK$ which increases as the temperature is reduced 
in the range $100\, mK<T<2\, K$. However, it is to be noted that the details of the behaviours are very different: Inductance noise acquires an approximate  $1/f$ dependence
only at the lowest temperatures, whilst at higher temperatures it is described by
much smaller exponents\cite{Anton2013}. Furthermore, at $f\sim1\, Hz$ the inductance
noise is almost temperature independent; at all frequencies the temperature
dependence ceases below $T\lesssim200\, mK$. Finally, the power saturation
of the noise seen in our resonators (fig~\ref{fig:PSD_power}) implies that the frequency of the relevant excitation
is close to the frequency of the resonator, e.g. $f\sim6\, GHz$. These
frequencies are too large for a dilute spin system with the densities
consistent with direct observations \cite{sendelbach2009,Anton2013}.
We have also performed measurements in weak ($\approx$ mT) in-plane fields which would be expected to affect these spin system, but not the superconductor\cite{healey2008magnetic}, and observe no change in the noise level.

Next we consider whether the STM could describe the jitter in the resonator frequency.  
The central assumption of the STM is that certain atoms or groups of atoms may tunnel between roughly equivalent potential 
energy minima separated in energy by an asymmetric splitting ${\Delta_{0}}$.
The tunnelling coupling energy is given by $\Delta\sim e^{-\lambda}$, where $\lambda$ is
a parameter describing the extent of wave-function overlapping between the states in the wells.
The Hamiltonian of each TLS has the form $H=\frac{1}{2}(\Delta_{0}\sigma^{z}+\Delta\sigma^{x})$,
where ${\sigma^i~(i=x,y,z)}$ are the Pauli matrices. ${E=\sqrt{\Delta_{0}^2+\Delta^2}}$ is the TLS energy splitting. The STM assumes that the
energies $\Delta_{0}$ and $\Delta$ are uncorrelated and have joint
probability distribution: $\displaystyle{P(\Delta_{0},\Delta)=\frac{\bar{P}_0}{\Delta}}$ in
a wide (atomic) range of energies, with a constant ${\bar{P}_0}$ which is material-dependent.

The TLS interact with the external electric field $\vec{{\cal E}}$ and the strain field $\varepsilon$ according to the Hamiltonian $\displaystyle{
H_{int}= \left [\vec{d}_0 \cdot \vec{{\cal E}} + \gamma \varepsilon \right ] \sigma^z}$
where ${\vec{d}_0}$ is the electric dipole moment and $\gamma$ is the elastic dipole moment. Different TLS interact between themselves
via long range dipole-dipole interaction (or phonon exchange) with
an effective strength given by the dimensionless parameter  ${\bar{P}_0 d_0^2/\epsilon ( \approx \bar{P}_0 \gamma^2/\nu v^2)}$ for electric (elastic) interactions, where $\epsilon$ is the dielectric constant of the medium, $\nu$ is the density of the glass and $v$ is the sound velocity. The dimensionless parameter is very small  ${\approx10^{-3} - 10^{-2}}$ such that within the STM  
interaction between TLS is assumed to be irrelevant. The dephasing and relaxation
of the TLS is due to their interaction with phonons; giving a TLS relaxation rate ${\Gamma_1 = [\gamma^2/2 \pi h^4 \nu v^5] \Delta^2 E \coth \left (E/2 k_B T\right )} $, where $k_B$ is the Boltzmann constant. The probability distribution of $\Gamma_1$ due to the distribution of $\Delta$ and $\Delta_{0}$ for a given $E$ is wide and increases very rapidly near ${\Gamma_1=0}$ \cite{black1977spectral}.  
Thermal fluctuations - absorption
or emission of thermal phonons by a collection of  independent TLS - 
could cause a jitter in the resonators centre frequency (equivalent to a fluctuating dielectric constant ${\delta \epsilon(r,t)}$).  
Since the number of activated TLS is proportional to temperature and the effect of each TLS is small in ${T/\nu_0}$, the STM  predicts\cite{hunklinger1976physical} a noise spectrum 
that vanishes quadratically for ${T \ll h \nu_0/2 K_B}$, in striking contrast to the results of fig~\ref{fig:Tem1}.  Moreover, in the regime of strong applied electric field our calculations show that one should expect a very weak power dependence of the noise spectra, in disagreement with the results of fig~\ref{fig:PSD_power}b. As such the STM is completely incompatible with our results. 

Other experimental studies have also found evidence for deviations from the STM. Individual TLS can be studied spectroscopically by superconducting qubits, allowing direct measurement of the relaxation time $T_{1}$ and the dephasing
time $T_{2}$ of \emph{individual} TLS\cite{shalibo2010lifetime}. The temperature dependence of these lifetimes has also been studied\cite{lisenfeld2010}, finding the expected
wide distribution of relaxation rates for TLS, but in disagreement with STM predictions the temperature dependence
of the relaxation rate is strong. Furthermore, the dephasing rate shows an anomalous behaviour: $\Gamma_{2} \sim T^{1.24}$. Studies of bulk ensembles of TLS using high-Q superconducting resonators have also found deviations from STM predictions with resonators showing a logarithmic power dependence in contrast to the square root dependence predicted by the STM\cite{Faoro2012,Lindstrom2009}. Secondly, the extracted  density of states of TLS  shows a weak power law dependence $P\propto E^{\mu}$, with $\mu=0.28$ \cite{Skacel2013pc}. 
This is in agreement with several studies in glasses where a slight suppression of the density of states at low energies: $\rho(E) \propto \Delta_{0}^{\mu}$ with $ \mu \approx 0.3$ was found \cite{zeller1971thermal} \cite{geva1997theory}. 
Finally, it was found that in thin oxide $a-SiO_{2+x}$ layers, dipole-dipole interactions between TLS may play a key role up to 100 mK \cite{ladieu2003dielectric}.

All of this leads us to conclude that at least for thin oxide layers
the STM does not  correctly describe the properties of TLS. Instead we argue that the noise originates from thermally activated
slow fluctuators that are also responsible for $1/f$ charge noise in superconducting
devices \cite{Kogan2008}.  In particular the resonator noise is generated by the switching of fluctuators that are located in the vicinity of
quantum TLS which are in resonance with the high frequency electric
field.  
In contrast to the STM, the interaction between TLS is relevant for our model and the density of low-energy states is slightly non-uniform.  In fact the non-uniform density of states is a natural consequence of strong interaction because the latter  always leads to the formation of the 
pseudogap by Efros-Shklovskii mechanism \cite{efros1975coulomb}. It is noted that the conclusion of
the small size of the interaction is based on the assumption that TLS
are local objects located at distances much larger than their size
which was not proven experimentally -- even for conventional glasses. 

We now show qualitatively that the assumptions of the dominant role
of TLS-TLS interaction and the energy dependent density of states explain
the present data as well as the results of  phase qubit experiments\cite{Grabovskij2012,lisenfeld2010}. 

Similarly to that of optical homogeneous linewidth \cite{lyo1982anomalous}, the dephasing rate of the resonant TLS can be estimated by associating the dephasing process with the thermal relaxation of off-resonant TLS that interact strongly with it. Since for a density of states ${\rho(E)\propto E^{\mu}}$, the number of the off-resonant TLS is proportional to ${T^{1+\mu}}$, the resulting width is $\Gamma_{2} \propto T^{1+\mu}$.  Due to this width, phononless relaxation processes between resonant TLS become possible with rate ${\Gamma_1 \propto T^{1+\mu}}$. Note that this power law is expected to hold at low temperature even whilst it dominates over the direct phonon emission.  From our data we obtain ${\mu \approx 0.2-0.4}$, as a result we get an estimate for the dephasing rate in very good agreement  with the one measured in phase qubit \cite{lisenfeld2010}, associated with a value of ${\mu =0.24}$. 
Furthermore the wide distribution of relaxation rates implies the presence of very slow TLS in the vicinity
of the resonant ones. These very slow TLS have coherence times much
shorter than the tunnelling rate and thus they can be considered as classical
fluctuators. Their presence is independently revealed by charge
fluctuations that display a $1/f$ power spectrum\cite{Kogan2008}. The effect
of the classical fluctuators on the resonant quantum TLS is to generate
an energy drift of the TLS with a  $1/f$ spectrum \cite{Faoro2012}.
However, this only occurs if the interaction strength between the classical and coherent TLS 
is greater than the TLS width, i.e. if the slow fluctuators
are located within a distance ${R_{max}=\left(\langle \alpha \rangle/\Gamma_{2}\right)^{1/3}}$
from the resonant  quantum TLS. An important consequence of this picture is that the noise increases as the temperature decreases. This is due to the fact that as the width decreases with temperature the energy drift becomes more visible. As a result, with decreasing temperature the noise {\em sensitivity} increases, even if the number of thermally excited fluctuators decreases. Analytical computations based on Bloch equations formalism give the following result  for the noise spectrum \cite{Faoro2013}:
\begin{equation}
S_{y}\propto\frac{1}{\sqrt{1+\Omega^{2}/\Gamma_{2}\Gamma_{1}}}\frac{1}{T^{1+\mu}}\frac{1}{f}\;
\label{eq1}
\end{equation}
where ${\Omega=2 \frac{\Delta}{E} \vec{d}_0 \cdot \vec{{\cal E}}}$ is the Rabi frequency and ${\vec{{\cal E}}}$ is the applied electric field.
Using the measured value of $\mu\approx 0.34$ (see fig. \ref{fig:Tem1}) allows this equation to describe all the results reported in this work. We emphasize the agreement of this value with that usually obtained for various glasses.

Our model can also account for the power dependence (see Fig~\ref{fig:PSD_power}b) . In fact, since ${\Gamma_1 \Gamma_2 \propto T^{2(1+\mu)}}$, the saturation term becomes temperature dependent.  For low powers the spectra has an overall strong temperature dependence proportional to ${T^{-(1+\mu)}}$, whereas at high power our theory predicts that the temperature dependence of the spectra is constant. Fig~\ref{fig:PSD_power}b is a  fit  to the overall temperature and power dependence  of the noise at $0.1$ Hz: $\displaystyle{S_y= \frac{A}{\sqrt{T^{2(1+\mu)}+ B W}}}$. In the fit, we used the measured value of $\mu\approx0.34$; where $A$ and $B$ are sample-dependent parameters. We find that $A \approx~(4)\text{x}10^{-16}$ for all temperatures. However, the value of $B$ appears to change  when the temperature is decreased below $\sim$~200~mK. For temperatures above 200~mK we find $B$~=~3.5x10$^{-3}$ and for temperatures below 200~mK we find $B$~=~2.5x10$^{-4}$.  As previously explained, the model is justified at lowest temperatures, so the observed cross-over between the two regimes is not surprising. 
It is worth noting that while saturation is seen at all temperatures in fig~\ref{fig:PSD_power}, the saturation at the lowest temperature is the least clear. Although it should be noted that clear saturation in the low temperature limit has been observed elsewhere\cite{neill2013}.

We note that the relevance of the interaction between TLS in thin oxide layers does not contradict  the existing body of the data obtained previously on bulk amorphous materials. However, we do put forward the notion that results within literature on bulk materials may also benefit from a model including a dominant TLS-TLS interaction.

In conclusion, we have performed measurements of the $1/f$ noise
in superconducting resonators and have found an  inverse temperature
dependence with $S_{y}\propto1/T^{1+\mu}$, with $\mu$~=~{0.2-0.4} in strong agreement with results on individual TLS in phase qubits\cite{lisenfeld2010}.
We also find a strong scaling power dependence of the noise with temperature. We have demonstrated that these results are incompatible with a presence of surface spins ruled out the STM by observing of a non vanishing temperature dependence and a temperature dependent saturation power. Instead, our data can be explained by assuming a relevant interaction between TLS. This leads to a suppression of density of low energy states, with a small exponent ${\mu \approx 0.3}$ that is in remarkable agreement with previous measurement on glasses, and those on individual TLS lifetimes.

\begin{acknowledgments}
The authors would like to acknowledge M. Kataoka for critical reading of the manuscript, M. Gustafsson, S. de Graaf, L. Ioffe,
S. Anton and J. Clarke for fruitful discussions. This work was supported
by the NMS Pathfinder program, the EPSRC and by ARO W911NF-09-1-0395.
\end{acknowledgments}

\section{Methods}
{\bf Samples.}
A 50~nm thick epitaxial Nb film with a 5~nm Pt capping layer was grown by laser ablation on r-plane sapphire $(1\bar{1}02)$
substrate in an ultra-high vacuum chamber with a base pressure of $10^{-9}$
Torr and at a growth temperature of 650$^{o}$C. The typical Nb films
grown in these conditions\cite{mikhailov1997} were single-crystalline
Nb $(100)$ with low surface roughness, 0.3-0.5~nm. The film was
patterned into an array of resonators coupled to a common feedline
on a 10x5~mm$^{2}$ chip \cite{Lindstrom2009,Burnett2013} using
optical lithography and dry etching in a 2:1 ratio $\text{SF}_{6}$/Ar
plasma. 
The resonators used in our experiments are of the lumped element type,
consisting of an inductive meander (inductance $\sim$5 nH) in parallel
with an interdigital capacitor, see fig. \ref{fig:resonator}a. They
have centre frequencies ($\nu_{0}$) in the 4-8 GHz band, and intrinsic
quality factors ($Q$) above $10^{5}$. 

\begin{table}[h]
\begin{tabular}{|c|c|c|c|c|}
\hline 
Sample  & $\nu_{0}$  & $Q_{l}$  & $Q_{i}$  & $F\tan\delta_{i}$ \tabularnewline
\hline 
\hline 
Res1  & 5.55 GHz  & 70000  & 250000  & $1.4\text{x}10^{-6}$\tabularnewline
\hline 
Res2  & 6.68 GHz  & 78000  & 370000  & $1.1\text{x}10^{-6}$\tabularnewline
\hline 
\end{tabular}
\caption{Device parameters for the resonators used in this work. $F$ is the filling factor. $Q_l$ and $Q_i$ denote respectively the loaded and internal quality factor.}
\label{tab:params} 
\end{table}

{\bf Measurements.}
A Pound frequency locked loop
(described in detail in ref. \cite{Lindstrom2011}) is used for all
measurements. The high sensitivity and high bandwidth of this technique
makes it suitable for a variety of resonance measurements \cite{de2013charge,degraaf2013near}
including noise measurements\cite{Burnett2013}.

The intrinsic loss tangent - proportional to the concentration and dipole moment of TLS - of the resonators $\tan\delta_{i}$ was
measured using the standard technique of fitting the $\nu_{0}$ vs.
T dependence \cite{Lindstrom2009}. For both samples we find $F\tan\delta_{i}$,
where $F$ is a filling factor of order 1, to be around 1x10$^{-6}$
indicating very low dielectric loss.

Additional epitaxial samples without the Pt capping layer
were found to exhibit similar parameters but with $Q_{i}\lesssim100000$
and $F\tan\delta_{i}>$2x10$^{-6}$ which is similar to previous work
using sputtered Nb\cite{Lindstrom2009}. The Pt capping layer
both increased $Q_{i}$ by a factor of 5 and significantly reduced
the dielectric loss tangent. This improvement is probably
due to the Pt minimizing the  intrinsic surface oxide layer
on the  niobium surface. As predicted by STM theory, the presence of TLS gives rise to temperature dependent resonant frequency shifts. However, the low loss
tangent in our samples meant that a $\pm$10~mK temperature fluctuation corresponds
to a 200~Hz frequency shift which is smaller than the flicker noise
level at -100~dBm drive power\cite{Burnett2013}. Such insensitivity
to temperature fluctuations is a stringent requirement for measuring
the temperature dependence of any slow process which could otherwise
be masked by temperature induced drifts.

All measurements were performed in a well-shielded dilution refrigerator equipped with a cryogenic HEMT amplifier and  heavily attenuated microwave lines. Our measurement setup has
been described elsewhere, see refs. \cite{Lindstrom2011,Burnett2013}.

{\bf Data analysis.}
The measured data was processed by
calculating the overlapping Allan deviation (ADEV) from the acquired time series . Each ADEV was then 
pre-screened for any signs of temperature drift, in which case the
dataset was discarded. By fitting to the $1/f$ part of the ADEV,
which for these sample occur for $t>0.01$s, we then parametrize the
frequency noise as $S_{y}\propto h_{-1}/f$, where $S_{y}$ is the
fractional frequency spectra (defined as $\left\langle \delta f(t)\delta f(t')\right\rangle /\nu_{0}^{2}$), and extract $h_{-1}$.  for plotting
the data in the more familiar form as the noise level at 0.1~Hz (calculated
from the $h_{-1}$ value).

\end{document}